\documentstyle[A4,prb,aps,graphicx,epsfig,floats,multicol]{revtex}

\begin{document}
\setlength{\baselineskip}{3ex}

\title{Sr$_2$V$_3$O$_9$ and Ba$_2$V$_3$O$_9$: quasi one-dimensional
spin-systems with an anomalous low temperature susceptibility }

\author{E. E. Kaul$^1$, H. Rosner$^2$, V. Yushankhai$^1$,
J. Sichelschmidt$^1$, R. V. Shpanchenko$^3$ and C. Geibel$^1$}

\address{$^1$Max Planck Institut f\"ur Chemische Physik fester Stoffe,
N\"othnitzer Str. 40, 01187 Dresden, Germany}
\address{$^2$Department of Physics, University of California, Shields
Avenue, Davis, CA 95616, USA}
\address{$^3$Department of Chemistry, Moscow State University, 199899
Moscow, Russia}

\date{\today}
\maketitle

\begin{abstract}

The magnetic behaviour of the low-dimensional Vanadium-oxides Sr$_2$V$_3$O$_9$ and
Ba$_2$V$_3$O$_9$ was investigated by means of magnetic susceptibility and specific
heat measurements. In both compounds, the results can be very well described by an
$S=\frac{1}{2}$ Heisenberg antiferromagnetic chain with an {\it intrachain} exchange
of $J = 82\ K$ and $J = 94\ K$ in Sr$_2$V$_3$O$_9$ and Ba$_2$V$_3$O$_9$,
respectively. In Sr$_2$V$_3$O$_9$, antiferromagnetic ordering at $T_N = 5.3\ K$
indicate a weak {\it interchain} exchange of the order of $J_{\perp} \cong 2\ K$. In
contrast, no evidence for magnetic order was found in Ba$_2$V$_3$O$_9$ down to 0.5 K,
pointing to an even smaller {\it interchain} coupling. In both compounds, we observe
a pronounced Curie-like increase of $\chi$(T) below 30 K, which we tentatively
attribute to a staggered field effect induced by the applied magnetic field. Results
of LDA calculations support the quasi one-dimensional character and indicate that in
Sr$_2$V$_3$O$_9$, the magnetic chain is perpendicular to the structural one with the
magnetic exchange being transferred through VO$_4$ tetrahedra.

PACS: 75.50.Ee, 75.40.Cx, 71.20.Ps

\end{abstract}

\pacs{75.50.Ee, 75.40.Cx, 71.20.Ps}


\section{Introduction}

The discovery of the first inorganic Spin-Peierls system CuGeO$_3$\cite{Hase1} has
lead to a strong revival of interest for one-dimensional spin systems. The thorough
investigation of this compound performed in the last years has revealed new
unexpected phenomena, like e.g. the coexistence of an antiferromagnetic state with
the Spin-Peierls state in slightly doped CuGeO$_3$\cite{Hase2}, as well as the
importance of frustration, i.e. antiferromagnetic exchange between second nearest
neighbours\cite{Fabricius}. Unfortunately, despite intensive research, no second
inorganic Spin-Peierls System, or example for a frustrated spin-chain could be yet
established. A further new topic which has emerged in this field in the past years is
the effect of a {\it Dzialoshinskii-Moriya}\cite{Dzial}$^,$\cite{Moriya} (DM)
interaction and  of a staggered g-factor anisotropy in such a spin-chain system. Both
theoretical and experimental studies indicate that they lead to a pronounced increase
of the susceptibility at low temperatures and to the opening of a gap in the magnetic
excitation spectra upon applying a magnetic
field\cite{Affleck1}$^,$\cite{Oshikawa}$^,$\cite{Feyerherm}.  However, there are
presently only very few systems where this effect could be investigated.

Most of the recent studies on one dimensional systems have been
carried out on compounds based on Cu$^{+2}$, where the magnetism
originates from the $S=\frac{1}{2}$, one {\it d}-hole
Cu$^{+2}$-state. Thorough investigations of one-dimensional systems
based on the electron counterparts, i.e. V$^{+4}$ or Ti$^{+3}$, are
scarce. For most of them only preliminary susceptibility measurements,
if at all, have been reported. This is surprising, since
Vanadium-based ternary and quaternary oxides form a large number of
compounds, many of them crystallizing in low dimensional
structures\cite{Ueda1}$^,$\cite{Ueda2}. Complex Vanadium oxides have a
very rich structural chemistry\cite{Schindler}, since Vanadium can be
coordinated in pyramidal, tetrahedral, or octahedral form (depending
on the oxidation state), and these polyhedra can be joined in a large
variety of ways\cite{Zavalij}$^,$\cite{Boudin}. Therefore, it is
possible to get interesting arrangements of magnetic cations,
resulting in anomalous magnetic properties. Thus e.g., unusual
magnetic behaviours were observed in the past years in a series of
compounds\cite{Korotin}$^,$\cite{Kanada}
(CaV$_2$O$_5$\cite{Onoda}$^,$\cite{Koo},
Ca(Sr)V$_3$O$_7$\cite{Liu}$^,$\cite{Whangbo}$^,$\cite{Takeo},
Ca(Sr)V$_4$O$_9$\cite{Pickett}$^,$\cite{Takano}$^,$\cite{Oka},
Pb$_2$V$_5$O$_{12}$\cite{Shpanchenko}) where V$^{+4}$ in pyramidal
coordination forms plaquettes connected in one or two dimensions. The
possibility to tune the Vanadium valence between +4 (one {\it d}
electron) and +3 (two {\it d} electrons) allows the realization of
both $S=\frac{1}{2}$ and $S = 1$ spin systems, where one expect the
strongest quantum effects. Compounds with an intermediate
valence-state, as for example
$\alpha$'-NaV$_2$O$_5$\cite{Isobe}$^,$\cite{Sawa}, allow to study the
interplay between magnetic and charge degrees of freedom. For both
V$^{+4}$ and V$^{+3}$ states, depending on the local configuration of
the V-atoms, degenerated or nearly degenerated $t_{2g}$ levels can be
achieved leading to additional orbital degrees of freedom. Combining
these degrees of freedom might lead to new interesting phenomena. Thus
V-compounds were e.g. suggested to be the ideal candidates for systems
with strong biquadratic
interactions\cite{Millet}$^,$\cite{Gavilano}$^,$\cite{Mila}.

We have therefore started a more detailed study of low dimensional
ternary and quaternary Vanadium oxides. We present here our
investigation of the magnetic and thermodynamic properties of
Sr$_2$V$_3$O$_9$ (SVO) and Ba$_2$V$_3$O$_9$ (BVO) and discuss our
experimental results using several basic models and band structure
calculations. Whereas in most of the low-dimensional vanadates
investigated so far V$^{+4}$ has a square pyramidal coordination, in
these two compounds V$^{+4}$ is located in an octahedral
environment. As our results show, this has a profound influence on the
magnetic properties.

\section{Structures}

\begin{figure}[!h]
\begin{minipage}[b]{1\linewidth}
\centering
 \includegraphics[width=2.4in]{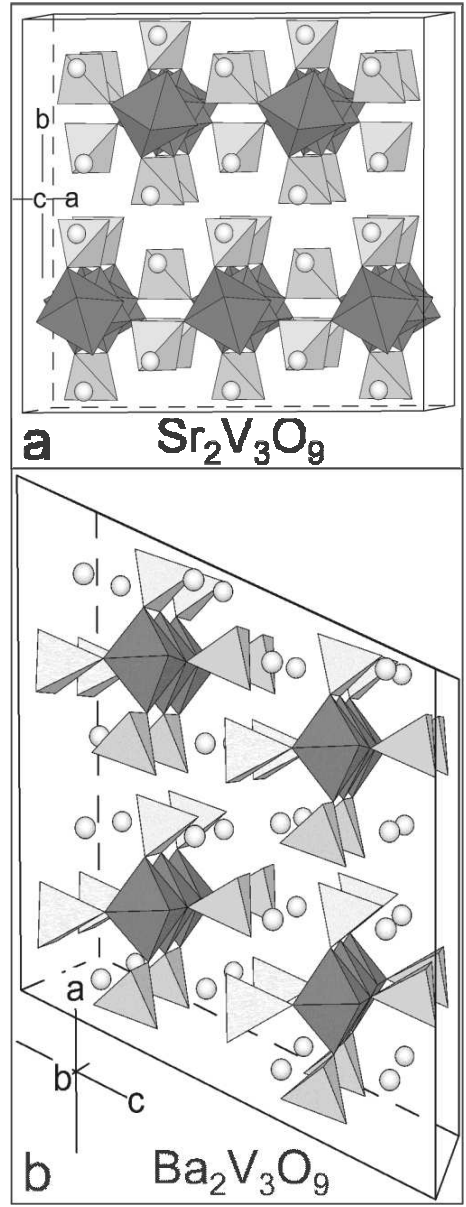}
\vspace{3mm}
  \caption{View of the crystal structures of Sr$_2$V$_3$O$_9$ (a) and Ba$_2$V$_3$O$_9$
(b) along the structural VO$_6$ octahedra chains. In this figure the V$^{+4}$O$_6$
octahedra are dark grey, the V$^{+5}$O$_4$ tetrahedra are light grey and the Sr or Ba
cations are shown as white spheres.}  \label{str1}
\end{minipage}
\end{figure}

SVO and BVO were first synthesized by Bouloux et al.\cite{Bouloux}. Despite the
identical stoichiometry, they crystallize in different structures, which were first
determined by J. Feldmann and Hk. M\"uller-Buschbaum \cite{Feldm SVO}$^,$\cite{Feldm
BVO} and latter on confirmed by O. Mentre et al.\cite{Mentre} and A. C. Dhaussy et
al. \cite{Dhaussy} (Fig.\ \ref{str1}). Both structures present three different
V-sites. Two of them show a tetrahedral oxygen-environment and can therefore be
assigned to V$^{+5}$, whereas one site shows an octahedral environment and can
therefore be assigned to V$^{+4}$. Thus, in contrast to the situation e.g. in
NaV$_{2}$O$_{5}$, there is a clear and complete V$^{+4}$ - V$^{+5}$ charge ordering.
The VO$_6$-octahedra form chains, which are arranged in a very different way in the
two compounds. In SVO, the octahedra are linked together by a common corner, whereas
in BVO, they are linked by a common edge. In SVO, two adjacent chains are connected
by VO$_4$-tetrahedra, leading to the formation of planes which are separated by
Sr-ions. In contrast, there is no simple connection between adjacent chains in the
Ba-based compound. Thus, BVO has a well defined one-dimensional structure whereas SVO
seems to be more two-dimensional.

\begin{figure}[!h]
\centering
 \includegraphics[width=2.4in]{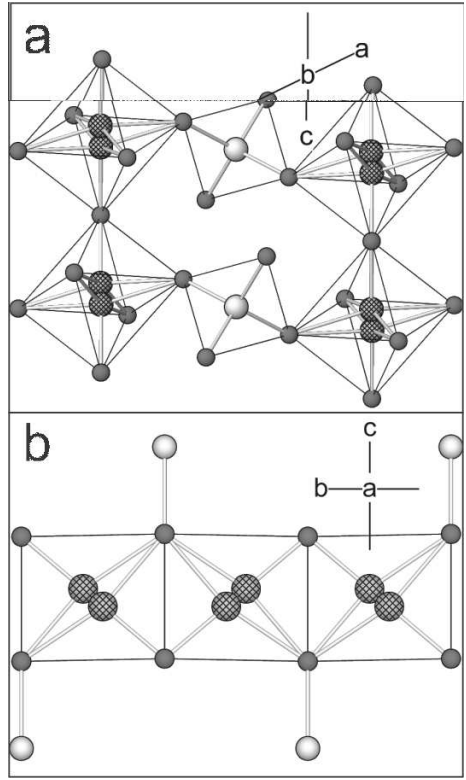}
\vspace{3mm}
\caption{Projection of the $VO_6$-Octahedra perpendicular to the chains
showing the splitting of the V-positions and how the octahedra are connected. a):
$Sr_2V_3O_9$; b) $Ba_2V_3O_9$. In this figure the big checked circles are V$^{+4}$,
the small dark grey ones are oxygen and the light grey ones are V$^{+5}$.}
\label{str2}
\end{figure}

In both compounds, the VO$_6$ octahedra are slightly distorted, and the V$^{+4}$-ion
is not located in the centre, but is slightly displaced by about $25\ pm$ from the
centre towards one of the O-ion (Fig.\ \ref{str2}). This result in the formation of
one short $V-O$ bond (the so called {\it vanadyl-bond}), with a $V-O$ distance $d
\approx 0.17 \ nm$, one opposite long $V-O$ bond, $d \approx 0.22\ nm$, and four
equatorial bonds of similar, average length $d \approx 0.20\ nm$ towards the O-atoms
forming the basal plane.			The presence of the vanadyl bond determines clearly the
{\it local axis} in each VO$_6$ octahedron. Normally the {\bf\^{z}} is taken parallel
to this short bond. The direction of the vanadyl-bond is suspected to be very
important for the magnetic properties, since it determines which {\it d} orbital is
occupied and therefore also determines the strength of the magnetic exchange along
different directions. According to simple crystal field considerations, the off
center displacement of the V-ion splits the $t_{2g}$ triplet into a low lying
$d_{xy}$-orbital singlet which is perpendicular to the vanadyl-bond and an excited
orbital doublet ($d_{xz}$\ and \ $d_{yz}$).

In SVO, the vanadyl bond is formed between the V$^{+4}$ and the O-ions
that connects the octahedra but it is not clear in wich sense it is
directed. The refinement of the XR-data\cite{Feldm SVO}$^,$\cite{Feldm
BVO} lead to a splitting of the V-site towards both directions, both
positions being statistically occupied to 50\%. From simple energy
considerations, one expect a long range ordering of the vanadyl bonds
within one chain, all pointing in the same sense, in order to avoid
one connecting O-ion being involved into two vanadyl bonds (in this
case the oxydation state of the Oxygen would be lower than -2). In
contrast, correlation in the direction of the vanadyl bond between
different chains might be absent because the difference in the
total-energy between parallel and antiparallel arrangement should be
extremely weak.  Thus the entire structure of the 3D ground state
remains unclear at present. In SVO the local axis of the octahedra are
tilted from the direction of the chains by an angle of $\sim 17.5^o$
and alternates along the chain. This alternation leads to a staggered
component of an anisotropic g-factor. At the same time, the occupied
{\it d}-orbital would be almost perpendicular to the
octahedra-chain. For that, one expect a very weak magnetic exchange
along the structural chain, because there is no overlap between the
occupied {\it d}-orbital and the {\it p}-orbitals of the connecting
{\it O}-atoms. On the contrary, one expect a much stronger magnetic
superexchange via the VO$_4$-tetrahedron connecting two VO$_6$
octahedra in adjacent chains. This would result in {\it magnetic}
spin-chains perpendicular to the structural octahedra-chains.

In BVO, with edge-sharing octahedra, the succesive planes of occupied d-orbitals and
the vanadyl bonds as its normal are arranged in a zig-zag manner along the structural
chain direction with an angle of $\sim46.5^o$ with respect to it. Whereas Feldmann
and M\"uller-Buschbaum claims a complete ordering of the vanadyl bonds, all bonds
pointing in the same sense\cite{Feldm BVO} , Dhaussy et al.\cite{Dhaussy} suggests
only ordering within each chain, all bonds pointing in the same sense within one
chain, but without correlation among the vanadyl bond between different chains. For
neighboring occupied d-orbitals a common edge O-ion provides a short superexchange
path via the {\it p} orbital lying parallel to the intersection of the vanadium
d-orbital planes. This leads to a significant magnetic exchange along the structural
chain.

Only little is known about the physical properties of both
compounds. O. Mentre et al.\cite{Mentre} and A.C. Dhaussy et
al.\cite{Dhaussy} performed preliminary measurements of the magnetic
susceptibility $\chi$(T). From the maximum observed in $\chi$(T) they
concluded that antiferromagnetic ordering occurs at $T_N = 50\ K$ and
$T_N = 58\ K$ in SVO and BVO, respectively. However, our results shall
show that this maximum is not related to a broadened antiferromagnetic
transition but to the onset of AF-correlations in the one-dimensional
spin-chains.

\section{Experimental Techniques}

Single phase polycrystalline samples of SVO and BVO were obtained by
the method of solid state reaction in dynamical high
vacuum. A$^{+2}$V$_3$O$_9$ (A=Sr, Ba) were synthesized from a
stoichiometric mixture of A$_2$V$_2$O$_7$ and VO$_2$ at 900 $^oC$ for
24 hs. A$_2$V$_2$O$_7$ were obtained heating a stoichiometric mixture
of ACO$_3$ and V$_2$O$_5$ in air in two steps of 24hs. each one at 850
and 900 $^oC$ respectively (with an intermediate ground). This
preparation procedure lead to dark-red brown (SVO) and light-red brown
(BVO) powder. Neither the crucible material (Al$_2$O$_3$ or Pt) nor
the form of the starting mixture (intimately mixed powder or pressed
pellet) had a significant influence on the results. The sample were
characterized using a STOE powder-diffractometer. For both compounds,
the diffractograms could be very well fitted using the structure
proposed by Feldmann\cite{Feldm SVO}$^,$\cite{Feldm BVO}. No extra
peaks corresponding to foreign phases were observed. The magnetic
susceptibility $\chi$(T) was measured between 2 and 400 K in fields up
to 5 Tesla on powder samples in a commercial (Quantum Design)
Squid. The specific heat ($C_p$) measurements were performed on
pressed pellets with a relaxation method using a commercial PPMS
equipment (Quantum Design). Measurements of the dielectric constants,
which are reported elsewhere\cite{Bobnar}, show that both compounds
are isolators.

\section{Quasi one-dimensional magnetic behaviour}

\begin{figure}[!h]
\centering
 \includegraphics[width=3.3in]{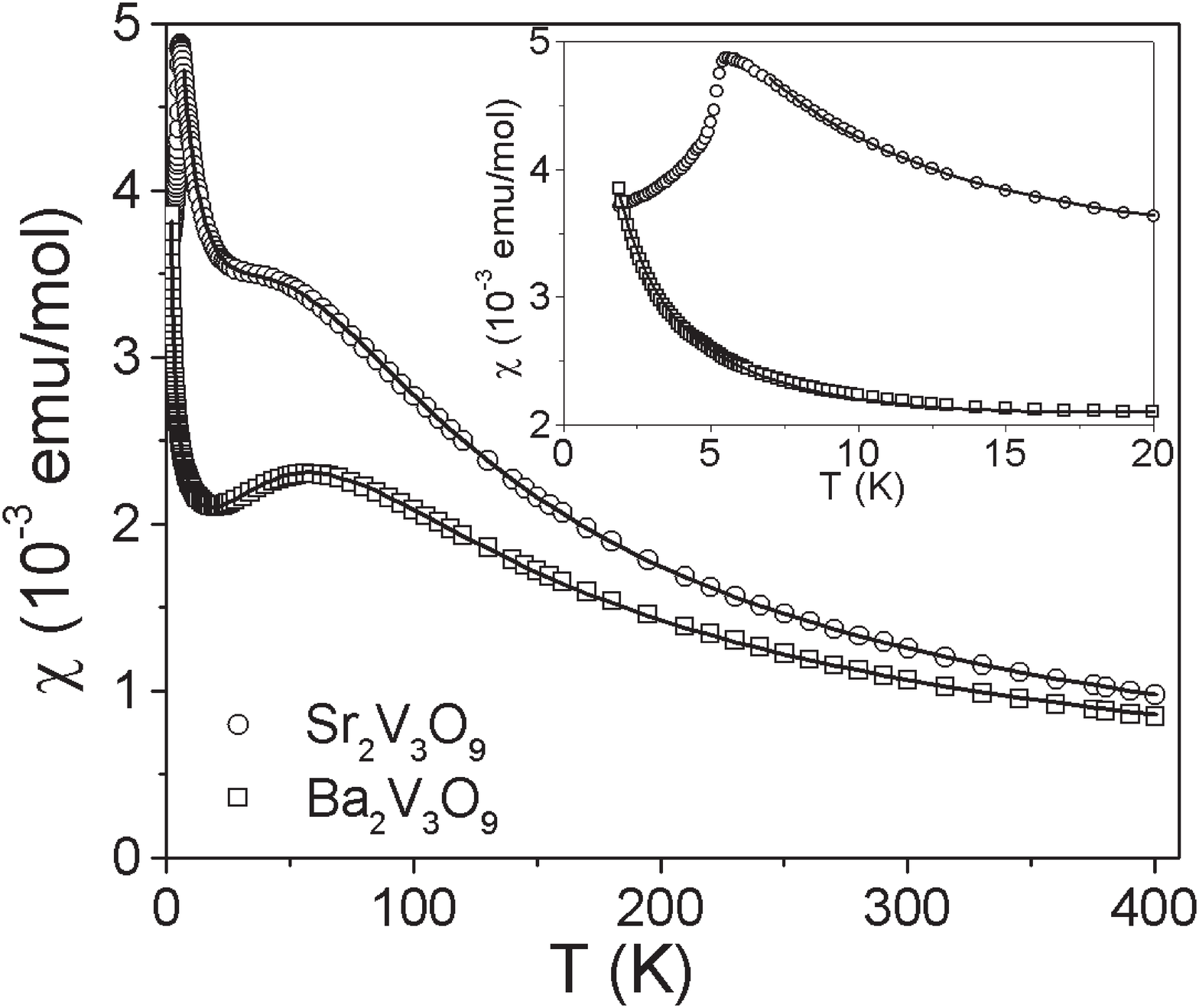}
 \vspace{3mm}
\caption{Temperature dependence of the susceptibility of SVO (circles) and BVO
(squares). The continuous lines show the fits of $\chi$(T) using equation 1 as
described in the text. Inset: susceptibility at low temperatures showing the
antiferromagnetic transition at 5.3 K in SVO and the absence of transition in BVO.}
\label{susc1}
\end{figure}

Despite the different crystallographic structures, the
susceptibilities of both compounds are rather similar (Fig.\
\ref{susc1}). $\chi$(T) follows a Curie-Weiss law at high
temperatures, pass through a maximum around $50\ K$ and $60\ K$ for
SVO and BVO, respectively, and increases again significantly below
$30\ K$. In SVO, a well defined anomaly is seen at $T_N = 5.3\ K$. The
pronounced decrease of $\chi$(T) below $T_N$ points to a transition
into an antiferromagnetic state. Since $\chi$(T) has still a rather
large value below $T_N$, a Spin-Peierls transition is unlikely. In
contrast, no anomaly is observed for BVO down to $2\ K$. Preliminary
ESR measurements show a $S=\frac{1}{2}$ spin susceptibility with a
temperature dependence almost identical to that of the bulk
susceptibility. The line shape observed in these measurements differs
from a Lorentzian, as expected for low dimensional spin systems.

The broad maximum in $\chi$(T) is a hallmark for low dimensional spin systems,
whereas the increase at low temperatures is rather unusual. We shall first focus on
the low-dimensional behaviour and discuss the low-temperature increase latter on. We
fit our susceptibility measurements with an expression of the form:

\begin{equation}
\chi(T)=\chi_{1D}(T)+\chi_{LT}(T)+\chi_{vv}(T) \label{suscfit}
\end{equation}

$\chi _{1D}$(T) is the contribution of the spin $S=\frac{1}{2}$
Heisenberg antiferromagnetic chains, which is known with a high
precision over the whole measured temperature range\cite{Eggert}. We
took the polynomial approximation of Feyerherm\cite{Feyerherm} (valid
for $T>0.05J/k_{B}$): \\
\begin{equation}
\chi_{1D}(T)=\frac{N_{A}\mu_{eff}^{2}}{3k_{B}T}
\frac{1+0.08516x+0.23351x^{2}}{1+0.73382x+0.13696x^{2}+0.53568x^{3}} \label{susc1D}
\end{equation}
\\
with $x=\frac{|J|}{K_BT}$.

The increase at low temperatures was accounted for by a Curie-Weiss
term, $\chi _{LT}$

\begin{equation}
\chi_{LT}(T)=\frac{C}{T+\theta}\label{suscLT}
\end{equation}

and we included a temperature-independent Van-Vleck contribution $\chi
_{vv}$.

With this approach, the experimental data could be fitted very nicely
in the range $7\ K < T < 400\ K$ for SVO and $2\ K < T < 400\ K$ for
BVO, respectively. The fits are shown as solid lines in Fig.\
\ref{susc1} and the fits parameters are given in Table\ \ref{Table
1}. The effective moments of $\mu_{eff} = 1,79\ \mu_B$ (SVO) and
$\mu_{eff} = 1,64\ \mu_B$ (BVO) obtained with these fits are close to
the spin-only value expected for V$^{+4}$ ions ($\mu_{eff} = 1,73\
\mu_B$).

The quality of the fits supports the picture of an antiferromagnetic
chain for both compounds. The values for the magnetic in-chain
exchange {\it J} we got from these fits, $J \cong 82\ K$ and $J \cong
94\ K$ for SVO and BVO, respectively, are similar to that obtained in
other V$^{+4}$ compounds. However, the susceptibility is not very
sensitive for discerning a {\it 1D} from a {\it 2D} spin-system since
the temperature dependence of $\chi$(T) is quite similar for a
$S=\frac{1}{2}$ chain and for a $S=\frac{1}{2}$ square lattice, and
differences in the absolute values can be accounted by changes in
$\mu_{eff}$. On the contrary, profound differences are expected in the
magnetic part of the specific heat at low temperatures. For a
$S=\frac{1}{2}$ chain, one expect a linear term, whereas for a square
lattice, one expect a leading quadratic term. Thus more insight can be
gained from the analysis of the specific heat.

\begin{table}[!h]

\begin{tabular}{lcc}

  & SVO& BVO \\\hline $T_N(K)$&5.3&$<$0.5\\$J(K)$ & 82 & 94 \\
$\chi_{vv}(emu/mol)$&$2\times10^{-5}$&$1\times10^{-4}$\\$\theta(K)$&-4.5&-0.35\\$C(emu
K/mol)$&$2.9\times10^{-2}$&$4.8\times10^{-3}$\\$\mu_{eff}(\mu_B)$&1.79&1.64

\end{tabular}
\vspace{3mm} \caption[Table 1]{Parameters from the fits of the susceptibility for SVO
and BVO shown in Fig.\ \ref{susc1}} \label{Table 1}
\end{table}

\begin{figure}[!h]
\centering
 \includegraphics[width=3.3in]{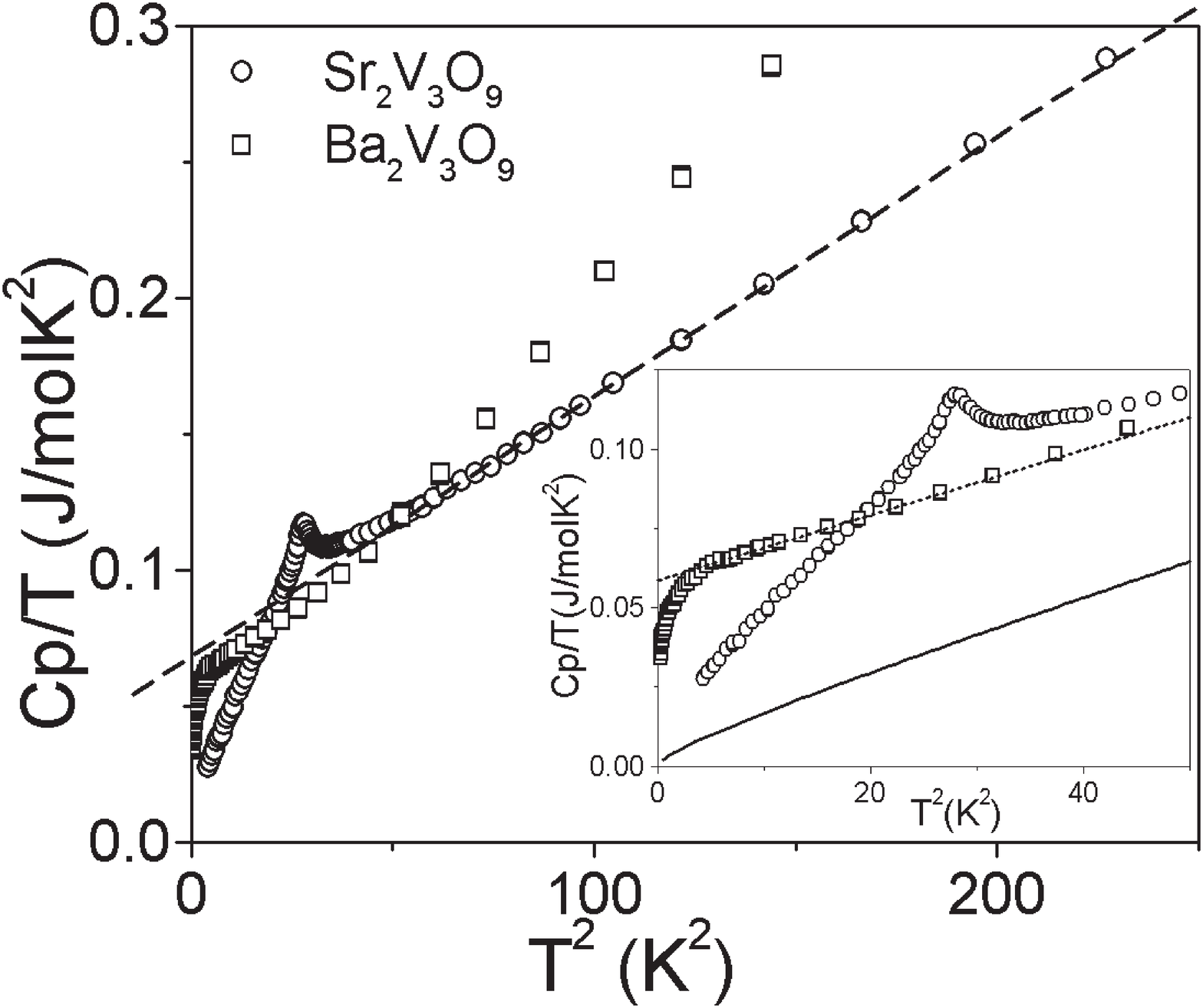}
 \vspace{3mm}
\caption{Specific heat of SVO (Circles) and BVO (squares) plotted as $C(T)/T$ versus
$T^2$. Inset: specific heat at low temperatures. The dashed and the dotted lines show
the fit $C(T) = \gamma T + \beta T^3$ as described in the text for SVO and BVO,
respectively. The solid line in the inset indicate the result expected for SVO using
a S=1/2 square lattice antiferromagnet model.} \label{Cp1}
\end{figure}

The results of the specific heat measurements in the temperature range $0.5\ K < T <
10\ K$ are shown for both compounds in Fig.\ \ref{Cp1} as $C_p/T$ versus $T^2$-plots,
in order to separate the different contributions. In both compounds, these plots
follow a straight line over a considerable temperature range, from $T_N$ to $15\ K$
in SVO and from $2.5\ K$ to $6\ K$ in BVO indicating that $C_p(T)$ is the sum of a
linear and a cubic contribution. Since the cubic term corresponds to the expected
contribution of the phonons, this demonstrate that the leading term of the magnetic
contribution in the unordered state (above $T_N$ for SVO) is linear in T. For a
$S=\frac{1}{2}$ antiferromagnetic Heisenberg-chain, theoretical
calculations\cite{Klumper}$^,$\cite{Johnston} predict for low temperatures ($T <
0.2J$):

\begin{equation}
\frac{C_p}{T} = \frac{2Rk_B}{3J} = \gamma_{theor} \label{Cp1Dteo}
\end{equation}

With the {\it J}-values obtained from the fit of $\chi$(T), this correspond to
predicted values $\gamma_{theor} = 68\ mJ/K^{2}mol$ (SVO) and $\gamma_{theor} = 59\
mJ/K^{2}mol$ (BVO). Fitting the specific heat data with:

\begin{equation}
C_p(T) = \gamma T + \beta T^3 \label{Cpgen}
\end{equation}

in the range $9\ K < T < 16\ K$ for SVO and $2.5\ K < T < 5\ K$ for
BVO, we obtained $\gamma = 66\ mJ/K^{2}mol$ (SVO) and $\gamma = 59\
mJ/K^{2}mol$ (BVO), very close to the values predicted from the
susceptibility results. This excellent agreement is a strong support
for the quasi one-dimensional nature of the spin system in both
compounds. In contrast, an attempt to explain the specific heat data
with a square lattice completely failed, as shall now be demonstrated
for SVO. For a two dimensional $S=\frac{1}{2}$ square array, the
leading term in the magnetic specific heat is expected to be
quadratic. We used the estimation of Takahashi\cite{Takahashi}:

\begin{equation}
C_p(T) = \delta T^2 = R\frac{3\zeta(3)}{4\pi J_{2D}^2m_{1}^{2}} T^2 \label{Cp2D}
\end{equation}

with $R$ the Gas constant, $\zeta(3)=1.202$ and
$m_1=1/2+0.078974$. $J_{2D}$ can be estimated from the temperature of
the maximum in $\chi$(T), $T_{\chi_{max}}$, since theoretical
calculations predict $T_{\chi_{max}} = 0.94 J_{2D}$. For SVO, this
lead to $J_{2D} = 56.5\ K$ and thus $\delta = 2.2 \times 10^{-3}\
J/K^{3}mol$. By adding the phonon contribution obtained from the $T^3$
term in the plot $C_p/T$ versus $T^2$ ($\beta=9.8\times10^{-4}\
J/K^3mol$), we get the total specific heat shown by the dashed lines
in Fig.\ \ref{Cp1}. At low temperatures, the calculated specific heat
is far below the experimental one, showing that even SVO is far from
being a square lattice. Thus the observation of a large linear
contribution in $C_p(T)$ at low temperatures, which matches very well
with that expected taking the {\it J}-values obtained from the
susceptibilities, is a very strong evidence that the magnetic coupling
in both compounds is predominantly one dimensional and that the
inter-chain coupling is weak. This inter-chain coupling $J_{\perp}$
can be estimated from the ordering temperature, since in weakly
coupled Heisenberg antiferromagnetic chains, $T_N$ is determined by
$J_{\perp}$. The exact relation is not known, but there exist a few
theoretical predictions. A relation which has been frequently used and
found to be quite reliable is that proposed by Schultz\cite{Schulz}
for an isotropic interchain coupling $J_{\perp}$:

\begin{equation}
|J_{\perp}| = \frac{T_{N}}{4 A\ ln\sqrt{(\frac{\Lambda J}{T_N})}}\label{Jperp}
\end{equation}

with $A = 0.32$ and $\Lambda \cong 5.8$. Using our experimental
results $T_N = 5.3\ K$ and $J = 82\ K$ for SVO, we obtain
$J_{\perp}\cong 1.9\ K$. For BVO, $T_N < 0.5\ K$ leads to $J_{\perp} <
0.15 K$. Thus the ratio between {\it intra}- and {\it inter}-chain
coupling is $2.3\times 10^{-2}$ in SVO and below $1.6\times 10^{-3}$
in BVO, showing that both compounds are quasi one-dimensional
spin-systems.

The specific heat of SVO shows at $T_N$ a small, but well defined anomaly. Further
on, there is a pronounced change in the $T$ dependence of $C_p(T)$ from above to
below $T_N$. Below $T_N$, the temperature dependence of $C_p(T)$ is dominated by a
large cubic term, which can be attributed to 3-dimensional magnons in the
antiferromagnetically ordered state. The weakness of the $C_p(T)$ anomaly at $T_N$ in
SVO is probably related to the small $\frac{T_N}{J}$ ratio. At these low temperatures
(compared to {\it J}), quantum fluctuations are very strong in quasi one-dimensional
spin-systems, reducing the size of the antiferromagnetically ordered moment and
accordingly the size of the specific heat anomaly. Preliminary measurements under
magnetic field indicate a broadening of this anomaly and a small shift to lower
temperatures in fields of a few Tesla. Thus the $C_p(T)$ results confirm an
antiferromagnetic phase transition at $T_N = 5.3\ K$ in SVO. In contrast, no anomaly
was seen in the $C_p(T)$ of BVO down to $0.5\ K$, although the slope also increases
towards low temperatures. The origin of this decrease is not clear yet.

\section{Low temperature $\chi$(T) upturn and annealing effects}

We now turn to the Curie-like upturn of $\chi$(T) at low temperatures
(LT-CW).  Usually, such an upturn is attributed to paramagnetic
moments due to impurities or defects in the sample. However, it was
recently demonstrated that such upturn can also be intrinsic. In quasi
one-dimensional $S=\frac{1}{2}$ systems it arises due to the staggered
field induced by the applied
field\cite{Affleck1}$^,$\cite{Oshikawa}$^,$\cite{Feyerherm}. The {\it
DM} interaction and a staggered g-factor anisotropy are two sources
for the staggered field effect. A first indication for the intrinsic
nature of the LT-CW in our compounds is given by the behaviour of
$\chi$(T) well below $T_N$ in SVO. If the LT-CW would be due to
paramagnetic impurities, it should also be present below
$T_N$. However, below $T_N$, $\chi$(T) decreases down to the lowest
measured temperature of 1.9 K, without any trace of paramagnetic
impurity contribution.

\begin{figure}[!h]
\centering
 \includegraphics[width=3.3in]{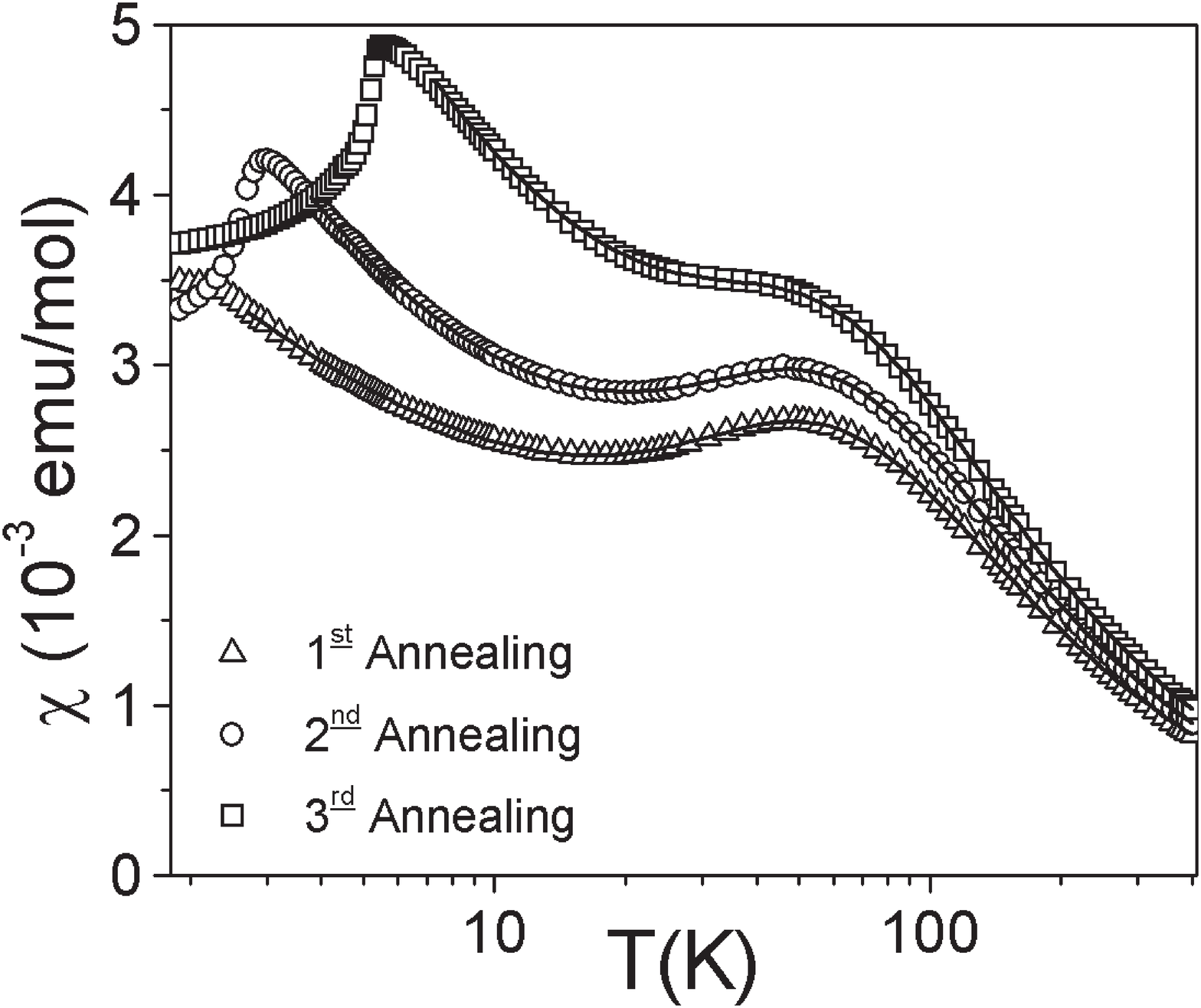}
 \vspace{3mm}
\caption{Susceptibility of SVO after successive annealing. The continuous lines show
the fits of $\chi$(T) using equation 1 as described in the text.}\label{susc2}
\end{figure}

\begin{figure}[!h]
\centering
 \includegraphics[width=3.3in]{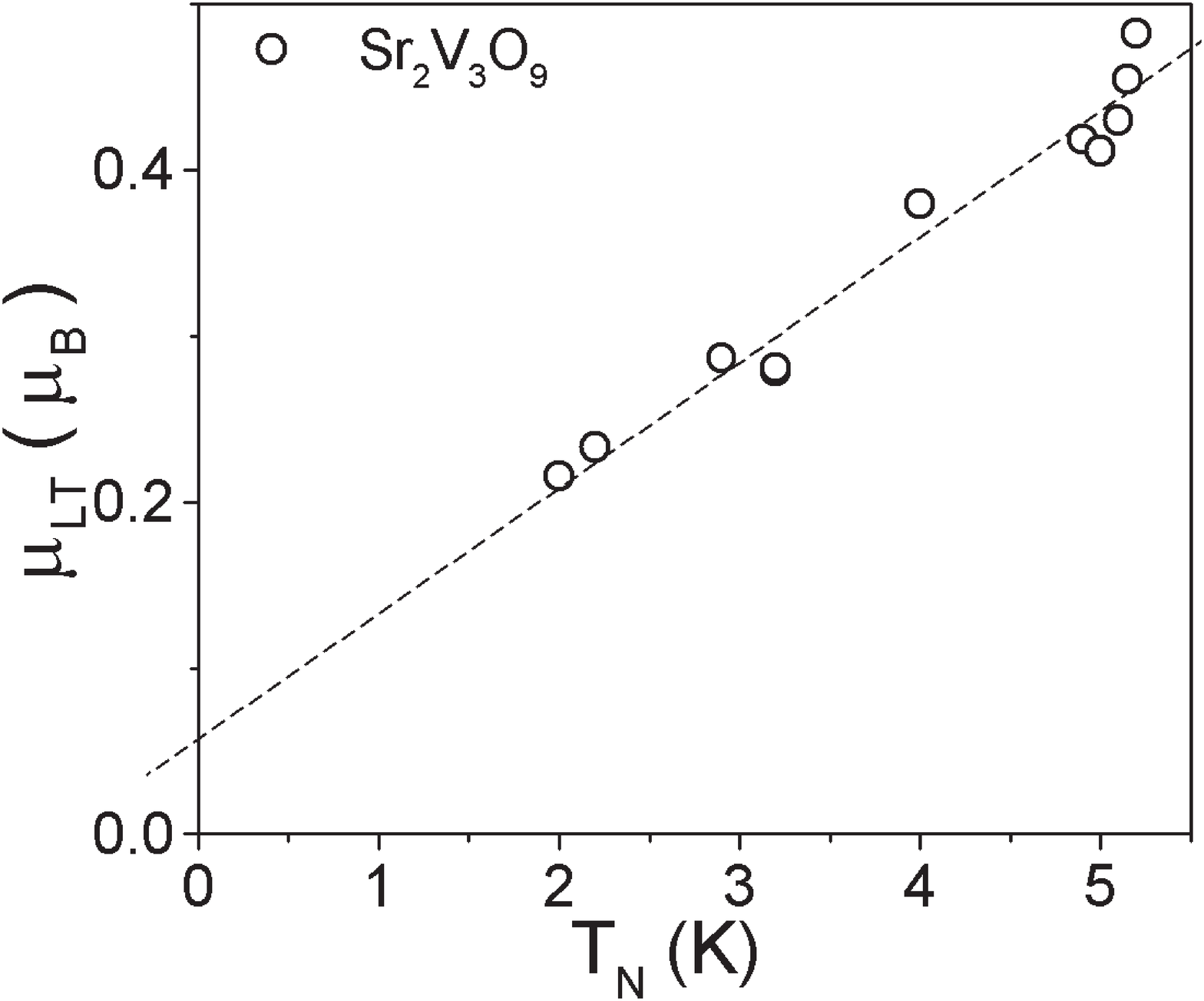}
 \vspace{3mm}
\caption{Correlation between the transition temperature into the antiferromagnetic
state $T_N$ and the effective moment $\mu_{LT}$ connected with the low temperature
increase in $\chi$(T) in SVO.}\label{Correl}
\end{figure}

In an attempt to get more insight into this phenomena, we investigated the influence
of different sample preparation conditions. We found that in BVO, $\chi$(T) was
almost insensitive to annealing, whereas in SVO, both the upturn in $\chi$(T) and the
antiferromagnetic ordering presented a clear and systematic dependence on the
annealing process. In Fig.\ \ref{susc2}, we show the evolution of $\chi$(T) of a SVO
sample after successive annealing steps at about 900 $^{o}C$ for 24 hrs in dynamical
vacuum. Whereas the changes at higher temperatures are rather weak, pronounced
differences are observed at low temperatures. The Curie-like upturn is always
present, but its magnitude increases with the number of annealing steps. Further on,
a strong shift of the transition temperature from $T_N = 2.4\ K$ in the {\it as
grown} sample up to 5.3\ K after the third annealing step is observed. Further
annealing did not increase $T_N$ any more suggesting that we reached saturation of
$T_N$. In order to perform a quantitative analysis, the $\chi$(T) curves were fitted
using Eq.\ (\ref{suscfit}) and the effective moment $\mu_{LT}$ connected with the
LT-CW was calculated from the coefficient {\it C}. All fit parameters for the three
curves of Fig.\ \ref{susc2} are given in Table\ \ref{Table 2}. $\mu_{LT}$ increases
by a factor of 2 when $T_N$ shifts from $2.4\ K$ to $5.3\ K$, whereas {\it J} does
not change appreciably and the high temperature effective moment increases only
slightly. Interestingly, we found an excellent but intriguing correlation between
$T_N$ and the $\mu_{LT}$, not only in this sample, but also in all investigated SVO
samples. This correlation is demonstrated in Fig.\ \ref{Correl}, where we plot $T_N$
versus $\mu_{LT}$ using all investigated samples. All points lies almost on one line.
Presently the origin of this linear relation is not clear. However, $\mu_{LT}$
increasing with $T_N$ is a further support that the LT-CW is not induced by defects.
Defects are expected to weaken a coherent three-dimensional magnetic coupling and
thus to reduce $T_N$. One would therefore expect a reduction of a defect induced
LT-CW with increasing $T_N$ which is opposite to what is observed. A high $T_N$
implies a well defined magnetic coupling between adjacent chains and hence one can
suspect that this requires a long range three dimensional order of the Vanadyl-bonds.
This is supported by the observation that high $T_N$s were observed in those samples
which were annealed as close as possible to the melting point, which is slightly
above 900 $^{o}C$. Since the ordering of the vanadyl-bonds between different chains
involve not the independent flip of single V-atoms, but the coherent rearrangement of
all the V-atoms within one octahedra-chain, one expect that this ordering can only
take place very close to the melting point. Unfortunately, an attempt to determine
changes in the ordering of the vanadyl bonds using Rietveld refinements of powder
X-Ray patterns failed, because the superstructure reflexes connected with this order
are too weak to be seen in powder patterns. These refinements showed some weak
differences in the structure, but it was impossible to get a significant information
about the ordering of the vanadyl bonds. One cannot exclude that some oxygen
vacancies arise during subsequential annealings. However, from our point of view it
looks unlikely since all oxygen atoms are connected either with vanadium V$^{+5}$
forming stable VO$^{-4}$ groups or involved in the formation of vanadyl bonds.

A correlation between a larger $T_N$ and a larger inter-chain coupling
can also be deduced from the specific heat. The specific heat at low
temperatures of the three sample whose susceptibility was shown in
Fig. 5 is plotted as $C_p/T$ versus $T^2$ in Fig. 7. The magnitude of
the linear term related to the one-dimensional spin-fluctuations is
almost the same in all samples, in accordance with the sample
independent {\it J}-value extracted from the analysis of $\chi$(T)
(Table \ref{Table 2}). This indicates that the intra-chain magnetic
exchange is not affected by the annealing process. However, the
temperature range where one can observe this linear term extends to
lower temperatures with decreasing $T_N$. This also shows that the
spin-fluctuations remains one-dimensional down to lower temperatures,
i.e that the inter-chain exchange is decreasing, in accordance with
the decrease of $T_N$. The absence of an upturn of $ C_p(T)$ at low
temperatures is an indication for the absence of paramagnetic
defects. One expect that a residual magnetic coupling of the defects
to the lattice should slightly lift the degeneracy of the magnetic
state of the defect and thus lead to a Shottky-like increase of
$C_p(T)$ towards low temperatures.

\begin{table}

\begin{tabular}{lccc}

  &$1^{st} Annealing$&$2^{nd} Annealing$&$3^{rd} Annealing$ \\\hline
$T_N(K)$&2.3&3&5.3\\ $J(K)$ & 81.3 & 80.8 & 82.2\\
$\chi_{vv}(emu/mol)$&$5\times10^{-5}$&$5\times10^{-5}$&$2\times10^{-5}$\\
$\theta(K)$&-2.3&-1.5&-4.5\\$C(emu
K/mol)$&$7.0\times10^{-3}$&$1.01\times10^{-2}$&$2.9\times10^{-2}$\\$\mu_{eff}(\mu_B)$&1.65&1.73&1.79

\end{tabular}
\vspace{3mm}
 \caption[Table 2]{Parameters from the fits of the susceptibility of SVO
shown in Fig.\ \ref{susc2}} \label{Table 2}
\end{table}

Thus, the analysis of the experimental results rules out that paramagnetic defects
are responsible for the increase of $\chi$(T) at low temperatures. Therefore, we have
to discuss possible intrinsic origins. The first trivial idea, that this upturn is
due to a weak ferromagnetic inter-chain interaction seems to be incorrect. The
inclusion of a weak ferromagnetic inter-chain coupling within a mean field approach
lead to a T-independent downwards shift of the $1/\chi(T)$ vs $T$ curve, i.e to a
larger increase of $\chi$(T) at $T_{\chi_{max}}$ than below $T_{\chi_{max}}$, in
contradiction to the experimental results. As stated above, the staggered field
effect caused by the DM-interaction and the staggered g-factor anisotropy ($g_s\sim
\pm(g_{\parallel}-g_{\perp})$) was recently demonstrated to be responsible for such
upturn in some one-dimensional Cu$^{+2}$ compounds\cite{Affleck1}$^,$\cite{Feyerherm}
and in $Yb_4As_3$\cite{Oshikawa}. Since our results are similar to the behaviour
found in these systems, the staggered field effect appears to be a promising
alternative explanation for our observations. According to Moriya\cite{Moriya}, a
DM-interaction is expected if the mid-point between two interacting magnetic ions is
not an inversion centre. Due to the low symmetry of our compounds, this condition is
fulfilled in both SVO and BVO.

Standard rules predicts the intensity of the DM interaction to be proportional to
$(\lambda / \Delta)J$, where $\lambda$ is the strength of the spin-orbit interaction
and $\Delta$ is the crystal field splitting of d-orbitals. One has $\lambda_V \simeq
0.03\ eV$ for V$^{+4}$ as compared to $\lambda_{Cu} \simeq 0.1\ eV$ for Cu$^{+2}$. At
the same time, the splitting of the $t_{2g}$ orbital levels due to the off-center
displacement of the V-ions ($\delta \simeq 0.2 $\ \AA) was estimated by us to give
$\Delta_V \simeq 0.2$ to $0.5\ eV$. In comparison, for Cu$^{+2}$ normally
$\Delta_{Cu} \simeq 1 $ to $ 2\ eV$. Preliminary ESR measurements performed in
SVO\cite{Ivanshin} indicate a rather small g-factor anisotropy with $g_s \sim
10^{-2}$. Therefore, one can expect that in the Cu-compounds\cite{Feyerherm}, where
the DM interaction and the $g_s$ contribution are nearly equally operative, the
resulting staggered field effect is somewhat stronger than that in the V-compounds
under consideration. Contrary, the low temperature Curie constants {\it C} measured
in BVO and, specially in SVO are even larger than the averaged one inferred from the
data reported for the {\it Cu} compounds. We remind, however, that in
Cu-benzoate\cite{Dender} the anomalous low temperature term $\chi_{LT}$ in $\chi$(T)
is also several times larger than that predicted in the theory\cite{Affleck1}.

With the DM-interaction as origin of the $\chi$(T) upturn, the
difference between the annealing effects in SVO and BVO can be
understood. As stated above, annealing should mostly affect the
relative orientation of the vanadyl bond between adjacent (structural)
V$^{+4}$-chains. Since in BVO, the structural V$^{+4}$ chain and the
magnetic chain are identical, a change of the orientation of the
vanadyl bond between adjacent chains shall not affect the magnetic
interaction along the magnetic chain.			 In SVO on the contrary, since
the magnetic and the structural chains are orthogonal, a change of the
orientation of the vanadyl bond between adjacent structural chains
shall change the symmetry between adjacent V$^{+4}$ ions along the
magnetic chain and thus should directly affect the strength of the
DM-interaction.

\begin{figure}[!h]
\centering
 \includegraphics[width=3.3in]{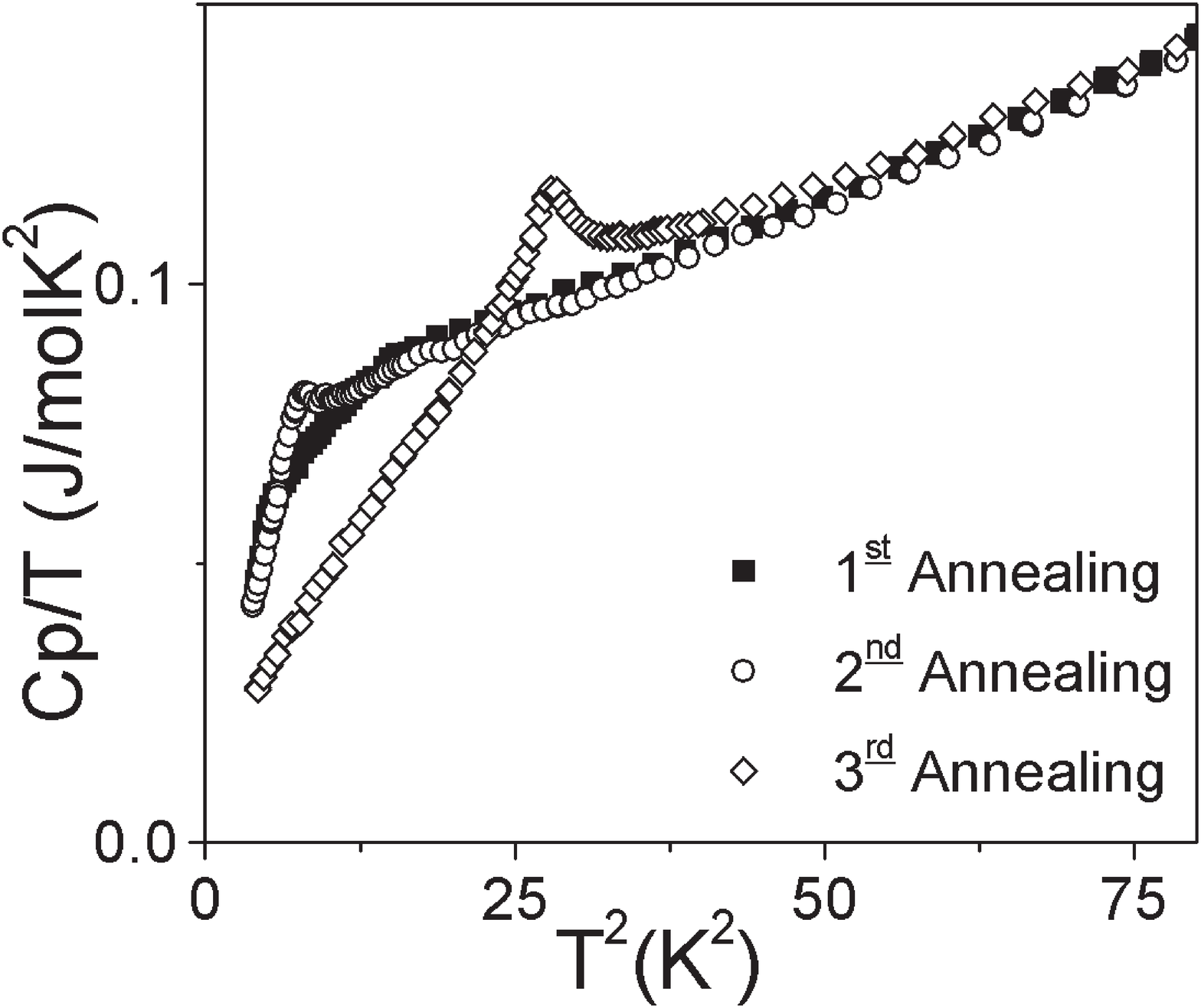}
 \vspace{3mm}
\caption{Specific heat of SVO after successive annealing. The samples used are the
same than in figure\ \ref{susc2}.}\label{Cp2}
\end{figure}

\section{Band structure calculations}

In order to get a better understanding and a consistent picture of the
properties of SVO and BVO, we performed electronic structure
calculations using the full-potential nonorthogonal local-orbital
minimum-basis scheme \cite{koepernik99} within the local density
approximation (LDA). In the scalar relativistic calculations we used
the exchange and correlation potential of Perdew and
Zunger\cite{perdew81}.			  V($3s$,$3p$,$4s$, $4p$, 3$d$), O(2$s$, 2$p$,
3$d$), Sr(4$s$, 4$p$, $5s$, $5p$, $4d$) and Ba($5s$, $5p$, $6s$, $6p$,
$5d$) states, respectively, were chosen as the basis set. All lower
lying states were treated as core states. The inclusion of
V(3$s$,3$p$) and Sr(4$s$, 4$p$) as well as Ba(5$s$, 5$p$) orbitals in
the valence states was necessary to account for non-negligible
core-core overlaps. The O 3$d$ states were taken into account to
increase the completeness of the basis set. The spatial extension of
the basis orbitals, controlled by a confining potential
\cite{eschrig89} $(r/r_0)^4$, was optimized to minimize the total
energy.

\begin{figure}[!h]
\centering
 \includegraphics[width=3.3in]{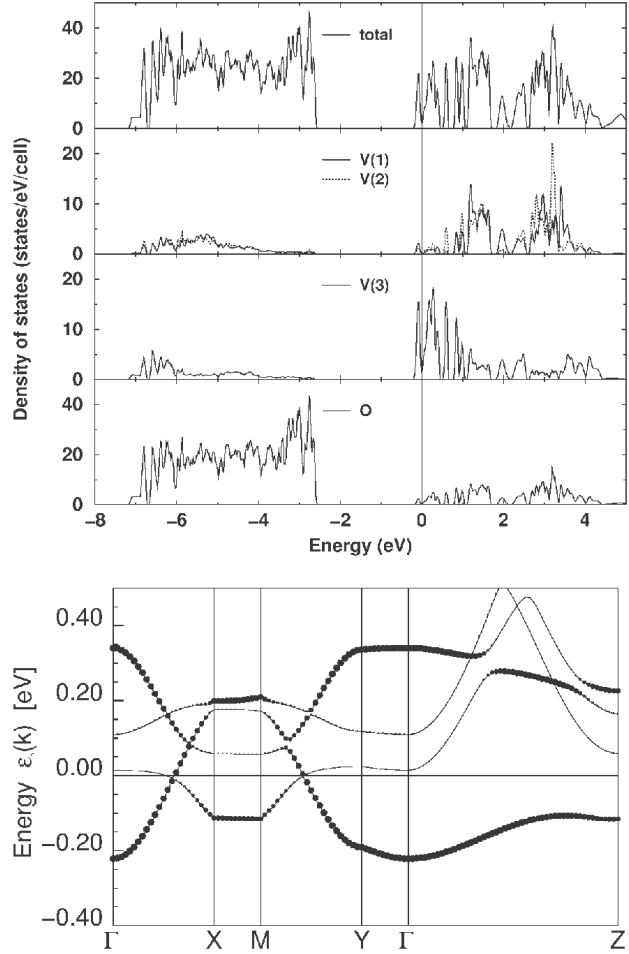}
\vspace{3mm} \caption{Upper panel: Total and partial density of states for
Sr$_2$V$_3$O$_9$. The Fermi level is at zero energy. Lower panel: The corresponding
band structure zoomed near the Fermi level. The size of the black circles illustrates
the contribution of the local V(3) 3d$_{xy}$ orbitals allowing a clear assignment of
the two half filled bands. The high symmetry points are noted as follows: $\Gamma$
(0,0,0), X (1,0,0), M (1,1,0), Y (0,1,0) and Z (0,0,1) in units of
(2$\pi/a$,2$\pi/b$,2$\pi/c$).}\label{fig_sr}
\end{figure}

For the sake of simplicity, for SVO, a crystal structure with vanadyl
bonds correlated between different chains (in phase) was calculated
using the space group $C1c1$ (No.~9). We assume that the influence of
the interchain correlation is of minor importance. The paramagnetic
calculations result in the total density of states (DOS) and the
partial DOS shown in the upper panel of Fig.~\ref{fig_sr}. We find a
valence band complex of about 7 eV width with two bands crossing the
Fermi level corresponding to the two formula units per unit
cell. Typical of vanadates, the valence band has mainly O 2$p$
character, with some admixture of V and small contributions from Sr
(not shown). The states at and right above the Fermi level are built
primarily from V 3$d$ orbitals, with the dispersion arising from
hybridization with the O 2$p$ states, with practically negligible
admixture of Sr states. In the partial DOS, we can clearly distinguish
between the two different types of V sites. The tetrahedrally
coordinated V(1) and V(2) have only a very small contribution at the
Fermi level.  Disregarding the spread overlap charge with the O 2$p$
valence states, this leads to a picture of V$^{+5}$ ions. On the other
hand, the octahedral V(3) site shows a half filled orbital at the
Fermi level, leading to a magnetically active spin 1/2 V$^{+4}$ ion.
Thus, the calculation is fully consistent with the empirical
assignment of the different V species mentioned earlier.

Analyzing the Slater-Koster-Integrals for the different V(3)-O terms
we find a much larger overlap for the short V(3)-O vanadyl bond
compared with the longer V(3)-O bond to the opposite corner of the
octahedron. Therefore, the electronic hopping along the octahedron
chain is suppressed. This can be clearly seen in the band structure
shown in the lower panel of Fig.~\ref{fig_sr}. The dispersion along
the ``structural'' chain direction $\Gamma$Z is rather weak and only
about 1/3 of the dispersion perpendicular to the ``magnetic'' chain
along the $\Gamma$X direction (corresponding to the crystallographic
$a$ direction). This fully confirms the conjecture based on an
empirical bond and orbital analysis made in section II. From the
viewpoint of the electronic structure, a picture of VO$_5$ pyramids
stacked along the $c$ axis seems more appropriate than the picture of
linked VO$_6$ octahedra. The dispersion in the third direction is
extremely small, leading therefore to spatially very anisotropic
exchange interactions for this compound.

In order to get a rough estimate for the exchange integrals, we analyze a tight
binding (TB) model taking into account as a first approximation nearest neighbors
(NN) only in each direction.  Taking the monoclinic angle into account, this results
in $t_a\sim$ 110 meV, $t_c\sim$ 30 meV and $t_b <$ 1meV. Here, $t_a$ corresponds to
the dispersion parallel to the ``magnetic chain'' running perpendicular to the $c$
direction of the ``structural'' chain (compare Fig.~\ref{str2}a). These transfer
integrals enable us to estimate the relevant exchange couplings, crucial for the
derivation and examination of magnetic model Hamiltonians of the spin-1/2 Heisenberg
type. In general, the total exchange $J$ can be divided into an antiferromagnetic and
a ferromagnetic contribution $J$ = $J^{AFM} + J^{FM}$.	   In the strongly correlated
limit, valid for typical vanadates, the former can be calculated in terms of the
one-band extended Hubbard model $J^{AFM}_{i}$ = $4t^2_{i}/(U_{eff})$. The index $i$
corresponds to NN in different directions, $U_{eff}$ is the effective on-site Coulomb
repulsion. Considering the fact that the VO$_5$ pyramids are not directly connected,
but only via VO$_4$ tetrahedra, ferromagnetic contributions $J^{FM}$ along the
``magnetic'' chain are expected to be small and we neglect them.\cite{remark_hr} From
LDA-DMFT(QMC) studies\cite{held01}, by fitting spectroscopic data to model
calculations\cite{mizokawa93} and similar LDA based model
calculations\cite{rosner02}, $U \sim 4$--$5$ eV is estimated for typical vanadates.
In rough approximation, this leads to exchange integrals of $J_a\sim$ 110 K,
$J_c\sim$ 10 K and $J_b <$ 0.1 mK. With respect to the crude approximations made, the
estimate for the leading exchange integral $J_a$ is in surprisingly good agreement
with the result $J_a$ = 82 K from the fit to the susceptibility data. At first
glance, the deviation from the experimentally estimated interchain exchange $J_\perp$
= 1.9 K looks rather large, therefore this problem will be discussed more detailed in
the following.

The (spatially) very anisotropic exchange interactions resembles
strongly the situation in the straight corner shared cuprate chains
Sr$_2$CuO$_3$ and Ca$_2$CuO$_3$. These quasi 1D cuprate compounds
exhibit antiferromagnetic intrachain exchange integrals J$_\parallel$
of the order of 2000 K, but they order antiferromagnetically at T$_N
\approx$ 5 K and T$_N \approx$ 9 K
only,\cite{kojima96,yamada95,kojima97} respectively, due to the very
weak interchain coupling. As in SVO, in those systems the coupling in
the weakest direction is negligibly small and will be dominated even
by the dipolar interactions. It has been shown\cite{rosner97} that for
such very anisotropic scenarios the application of Schulz's quantum
spin-chain approach\cite{Schulz} assuming isotropic interchain
interactions $J_\perp$, is limited. Assuming $J_\perp$ = 1/2 ($J_b$ +
$J_c$), the Ne\'el temperatures for the quasi 1D-cuprate chains were
considerably overestimated in Ref.~\onlinecite{rosner97}. Most
remarkably, an empirical estimate of $J_\perp$ in Sr$_2$CuO$_3$ in the
same work,\cite{rosner97} using the experimentally observed Ne\'el
temperature, leads to a value for the (isotropic) interchain exchange
smaller by a factor of about 2--3 than the theoretical estimate. This
is very much in line with the present scenario ($J_\perp^{theor}$ =
1/2 ($J_b$ + $J_c$) = 5 K and $J_\perp^{emp}$ = 1.9 K) and a strong
indication that the magnetic coupling in such anisotropic systems
deserves further theoretical investigation. Another source of the
discrepancy could originate from the remaining structural disorder,
which obviously has a strong influence on the ordering temperature
(see Tab. \ref{Table 2}).

\begin{figure}[!h]
\centering
 \includegraphics[width=3.3in]{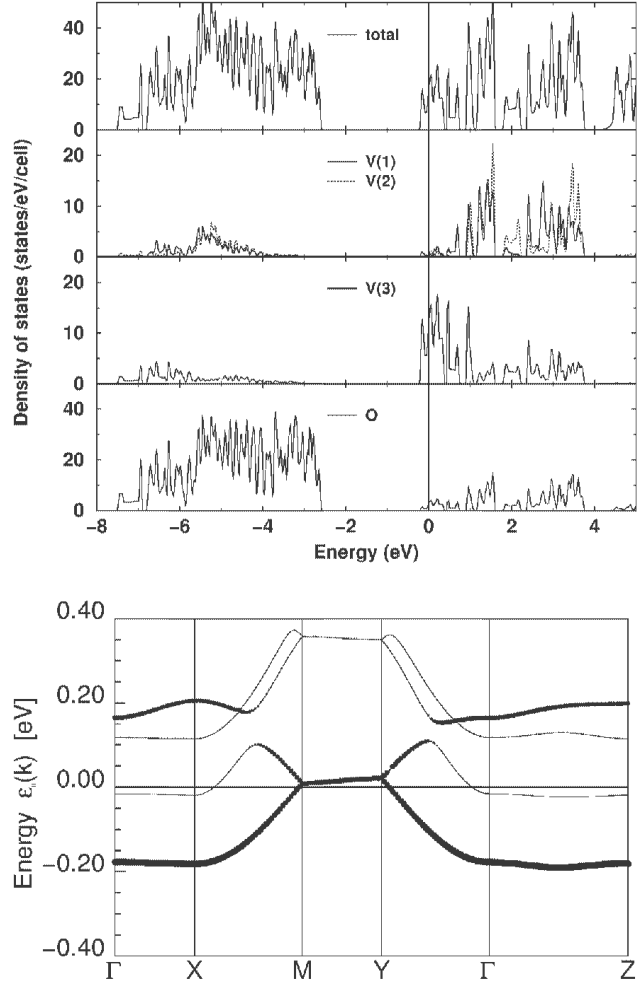}
 \vspace{3mm}
\caption{The same as in Fig.\ref{fig_sr} but for Ba$_2$V$_3$O$_9$.}\label{fig_ba}
\end{figure}

Now, we turn to the BVO system. In general, the electronic structure
of BVO looks rather similar to SVO. The total and the partial DOS as
well as the band structure for BVO are shown in
Fig.~\ref{fig_ba}. Like in SVO, two vanadium species are clearly
distinct, leading to tetrahedrally coordinated nonmagnetic V$^{+5}$
ions (V(1) and V(2)) and a spin 1/2 V$^{+4}$ ion (V(3)) inside the
edge sharing octahedra. In contrast to SVO, the largest dispersion of
the bands at the Fermi level along the XM direction (see lower panel
in Fig.~\ref{fig_ba}) is parallel to the octahedra chain, confirming
the picture from the bond analysis in Section II. A formal TB analysis
in terms of NN interactions only like in SVO yields $t_b\sim$ 90 meV,
$t_a\sim$ 15 meV and $t_c\sim$ 15 meV. Analogous to the procedure in
SVO described above, this results in exchange couplings $J_a\sim$ 80 K
and $J_b = J_c \sim$ 2K. Again, the largest calculated coupling $J_a$
= 80 K is in good agreement with the value of 94 K from the fit to the
experimental data. A closer look at the band structure with respect to
the interchain couplings (lines $\Gamma$X, MY and $\Gamma$Z in
Fig.~\ref{fig_ba}) shows that an TB analysis in terms of NN only is
rather problematic. The strong non cosine-like shape of these
dispersions, especially along $\Gamma$Z, is due to contributions from
more distant neighbors. The estimate of the corresponding exchange
terms and the investigation of the resulting complicated interplay of
competing frustrating exchange interactions and their influence on the
ordering temperature is far beyond the scope of the present
work. Thus, the understanding of the absence of long range order in
BVO down to 0.5 K remains an open problem for further work.

It should be noted that in both systems, the crystal field splitting
$\Delta$ for the V(3) $3d$-$t_{2g}$ levels is unusual small in the
present band structure calculation compared to other vanadates. This
is a possible indication for an enhanced DM interaction (being
proportional to 1/$\Delta$). A quantitative estimate of the split
$\Delta$ from our calculation would be problematic due to the known
failure of LDA resulting in a metallic solution with the half filled
states at the Fermi level instead of an insulating behavior as
observed in the experiment. The calculation of $\Delta$,  using a more
appropriate method, e.g. LDA+$U$, is a future task.

\section {Summary}

We have investigated the magnetic and the thermodynamic properties of
two ternary Vanadium oxide compounds, SVO and BVO, where the magnetic
V$^{+4}$ ions are located off-centered in VO$_6$ octahedra, forming
chains. Analysis of the results show that both compounds are quasi
one-dimensional spin systems, with {\it intra-chain} exchanges $J =
82\ K$ and $J = 94\ K$ for SVO and BVO, respectively. In SVO,
antiferromagnetic ordering at $T_N = 5.3\ K$ indicate a weak {\it
inter-chain} exchange $J_{\perp} = 1.9\ K$. In contrast, no evidence
for magnetic order could be observed in BVO, pointing to a very small
$J_{\perp}$ ($< 0.15\ K$). The conjecture, based on simple standard
arguments, that in SVO the magnetic chain is perpendicular to the
structural VO$_6$-octahedra chain is confirmed by an analysis of
elaborated LDA calculations. Both compounds present an unusual upturn
of the susceptibility at low temperatures, which we suggest to be due
to a sizeable DM-interaction. In SVO the size of this low temperature
upturn as well as the antiferromagnetic ordering temperature increase
when the samples are annealed. This peculiar low temperature behaviour
shall be the subject of future detailed investigations carried out on
single crystals.


\begin{references}
\bibitem{remark_hr} For the direction of the ``structural'' chains, a
larger ferromagnetic contribution to $J$ can not be excluded, resulting in a somewhat
reduced total AFM exchange along this direction.
\end{references}
\end{document}